\documentclass{appolb}
\usepackage{epsfig}
\begin{document}
\eqsec
\title{STUDY OF THE LOW ENERGY DYNAMICS\\ 
IN THE $ppK^{+}K^{-}$ SYSTEM WITH THE COSY-11\\
 MAGNETIC SPECTROMETER
\thanks{Presented at the Symposium on Meson Physics, Cracow, 01-04 October 2008.}%
}
\author{Micha{\l}~Silarski, Pawe{\l}~Moskal, Damian~Gil, Jerzy~Smyrski\\
 on behalf of the COSY-11 collaboration
\address{\begin{center}Institute of Physics, Jagiellonian University, PL-30-059 Cracow, Poland\\
                 \&\\
                 Institute for Nuclear Physics and J{\"u}lich Center for Hadron Physics,\\
                 Research Center J{\"u}lich, D-52425 J{\"u}lich, Germany\end{center}
}}
\maketitle
\begin{abstract} 
The near threshold production of $K^{+}K^{-}$ pairs in proton-proton 
collisions has been investigated at the cooler synchrotron COSY below 
and above the threshold for the $\phi$ meson production.
The experimental excitation function determined for the $pp\rightarrow ppK^{+}K^{-}$ 
reaction differs from theoretical expectations including proton-proton final state 
interaction. The discrepancy may be assigned to the influence of $K^{+}K^{-}$ or $pK^{-}$ 
interaction. In this article
we present distributions of the cross section for the $pp\rightarrow ppK^{+}K^{-}$ 
reaction as a function of the 
invariant masses of two and three particle subsystems at excess 
energies of Q~=~10~MeV and 28~MeV.
\end{abstract}
\section{Introduction}
The basic motivation for investigating the $pp\to ppK^+K^-$ reaction near the kinematical 
threshold was comprehensively reviewed by Oelert at the very first Cracow Workshop on Meson Production
and Interaction~\cite{oelert}.
The main reason for such studies is
an attempt to understand the nature of scalar resonances
$f_{0}$(980) and $a_{0}$(980), whose masses are very close to the 
sum of $K^{+}$ and $K^{-}$ masses. Besides the standard 
interpretation as $q\bar{q}$ mesons~\cite{Morgan}, these 
resonances were also proposed to be $qq\bar{q}\bar{q}$ 
states~\cite{Jaffe}, $K\bar{K}$ molecules~\cite{Lohse, Weinstein}, 
hybrid $q\bar{q}$-meson-meson systems~\cite{Beveren} or even quark-less 
gluonic hadrons ~\cite{Johnson}. 
The strength of the $K\bar{K}$ interaction is a crucial 
quantity 
regarding the formation of a 
$K\bar{K}$ molecule, 
whereas
the $KN$ interaction is of importance
in view of the vigorous discussion 
concerning the structure of the excited hyperon $\Lambda$(1405) which is 
considered as a three quark system or as a $KN$ molecular state~\cite{Kaiser}.
Additionally, these interactions appear to be very important also with respect to other
physical
phenomena, like for example a modification of the neutron star properties 
due to possible kaon condensation~\cite{Li} or
properties of strange particles immersed in the dense 
nuclear medium studied by means of heavy ion collisions~\cite{Senger,Laue,Barth,Menzel}.
In our approach~\cite{fipsi,gil,Moskal} we endouver to learn about 
the $K^+K^-$ and $Kp$ interactions from the excitation function 
and from invariant mass distributions of cross sections for the $pp\to pp K^+K^-$ reaction.
\section{Excitation function for the near threshold $K^{+}K^{-}$ production}
\begin{figure}
\centering 
\includegraphics[width=0.7\textwidth,angle=0]{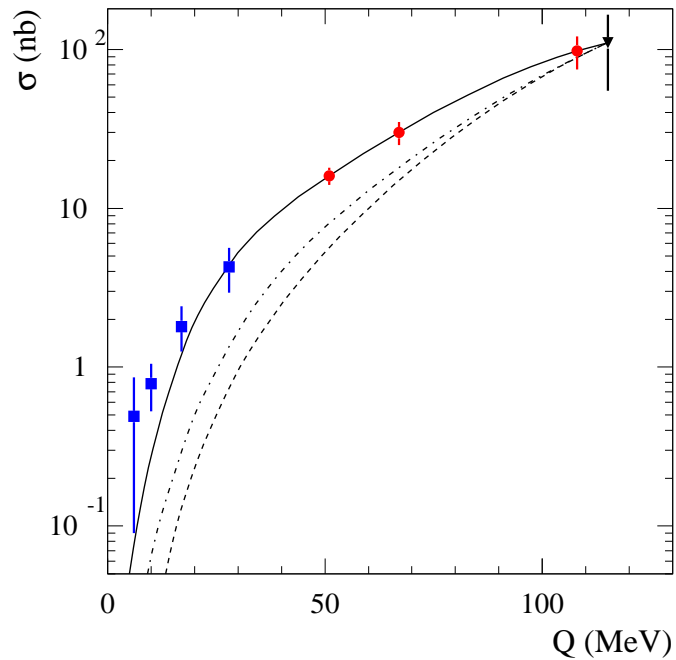}
\caption{Total cross section as a function of the excess energy Q for the reaction $pp\rightarrow ppK^{+}K^{-}$.
Triangle and circles represent the DISTO and ANKE measurements respectively.
The four points closest to the threshold are the results from COSY-11 measurements.
The curves are described in the text.}
\label{excitation-f}
\end{figure}

The measurements of the $pp\rightarrow ppK^{+}K^{-}$ reaction were 
conducted at low excess energies Q by the
collaborations ANKE~\cite{anke}, COSY-11~\cite{Quentmeier,Winter,Wolke} and 
DISTO~\cite{Balestra}. The achieved results are presented in Fig.~\ref{excitation-f} 
together with curves representing three different theoretical expectations~\cite{anke} 
normalized to the DISTO data point at Q~=~114~MeV.

The dashed curve represents the energy dependence from four-body phase space, when we assume 
that there is no interaction between particles in the final state. These calculations differ from 
the experimental data by two orders of magnitude at Q~=~10~MeV and by a factor of about five 
at Q~=~28~MeV. Hence, it is obvious, that the final state interaction effects in the 
$ppK^{+}K^{-}$ system cannot be neglected~\cite{oelert1}. 
 
Inclusion of the $pp$--FSI (dashed-dotted line in
Fig.~\ref{excitation-f}), by folding 
its parameterization known from the three body final state~\cite{pp-FSI} with the four 
body phase space, is closer to the experimental results, but does not fully 
account for the difference~\cite{Winter}. The enhancement may be due to the influence of 
$K^{+}K^{-}$ or $pK$ interaction which was neglected in the calculations. Indeed, as shown 
by authors of reference~\cite{anke,Colin_AIP}
the inclusion of the $pK^{-}$--FSI (solid line) reproduces the experimental data for 
the excess energies down to the point at Q~=~28~MeV. These calculations were accomplished 
under the assumption that the overall 
enhancement factor, originating from final state interaction in $ppK^{+}K^{-}$ system, 
can be factorised into enhancements in the $pp$ and two $pK^{-}$ subsystems~\cite{anke}:
 \begin{equation}
F_{FSI}~=~F_{pp}(q)\cdot F_{p_{1}K^{-}}(k_{1})\cdot F_{p_{2}K^{-}}(k_{2})~,
\label{pp-pkfsi}
\end{equation}
where $k_1$, $k_2$ and $q$ stands for relative momenta of particles in the first $pK^-$ subsystem,
second $pK^-$ subsystem and $pp$ subsystem, respectively. The factors describing the enhancement 
originating from $pK^{-}$--FSI are parametrised using the scattering length approximation:
\begin{equation}
F_{p_{i}K^{-}}~=~\frac{1}{1~-~i~k_{i}~a_{pK^{-}}}~,~~~~~~~~i=1,2
\label{MK-p}
\end{equation}
where $a_{pK^{-}}$ is a complex parameter describing the interaction, called effective 
scattering length. 
It is important to note that the inclusion of the $pp$ and $pK^{-}$ final state interaction 
is not sufficient to describe the data very close to threshold (see Fig.~\ref{excitation-f}).
This enhancement may be due to the influence of the $K^{+}K^{-}$ 
interaction, which was neglected in the calculations\footnote
{
It is worth mentioning, that in the calculations also the $pK^{+}$ 
interaction was neglected. It is repulsive and weak and hence it can be 
interpreted as an additional attraction in the $pK^{-}$ system~\cite{anke}.
}.
\newpage 
\section{The differential observables for COSY-11 data \\
 measured at Q~=~10~MeV and Q = 28 MeV}
\begin{figure}[h]
\centering
\includegraphics[width=0.3\textwidth,angle=0]{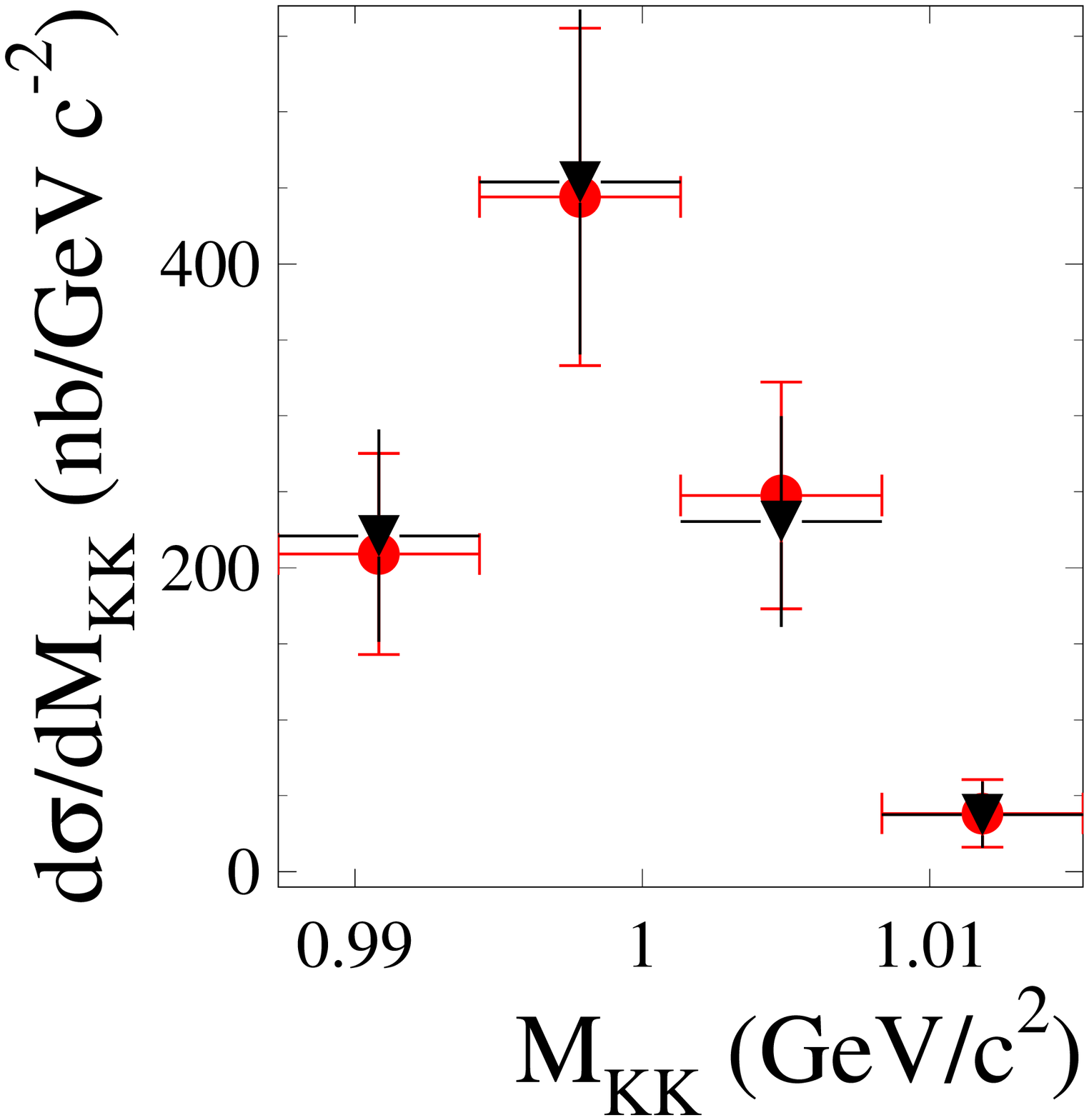}
\includegraphics[width=0.296\textwidth,angle=0]{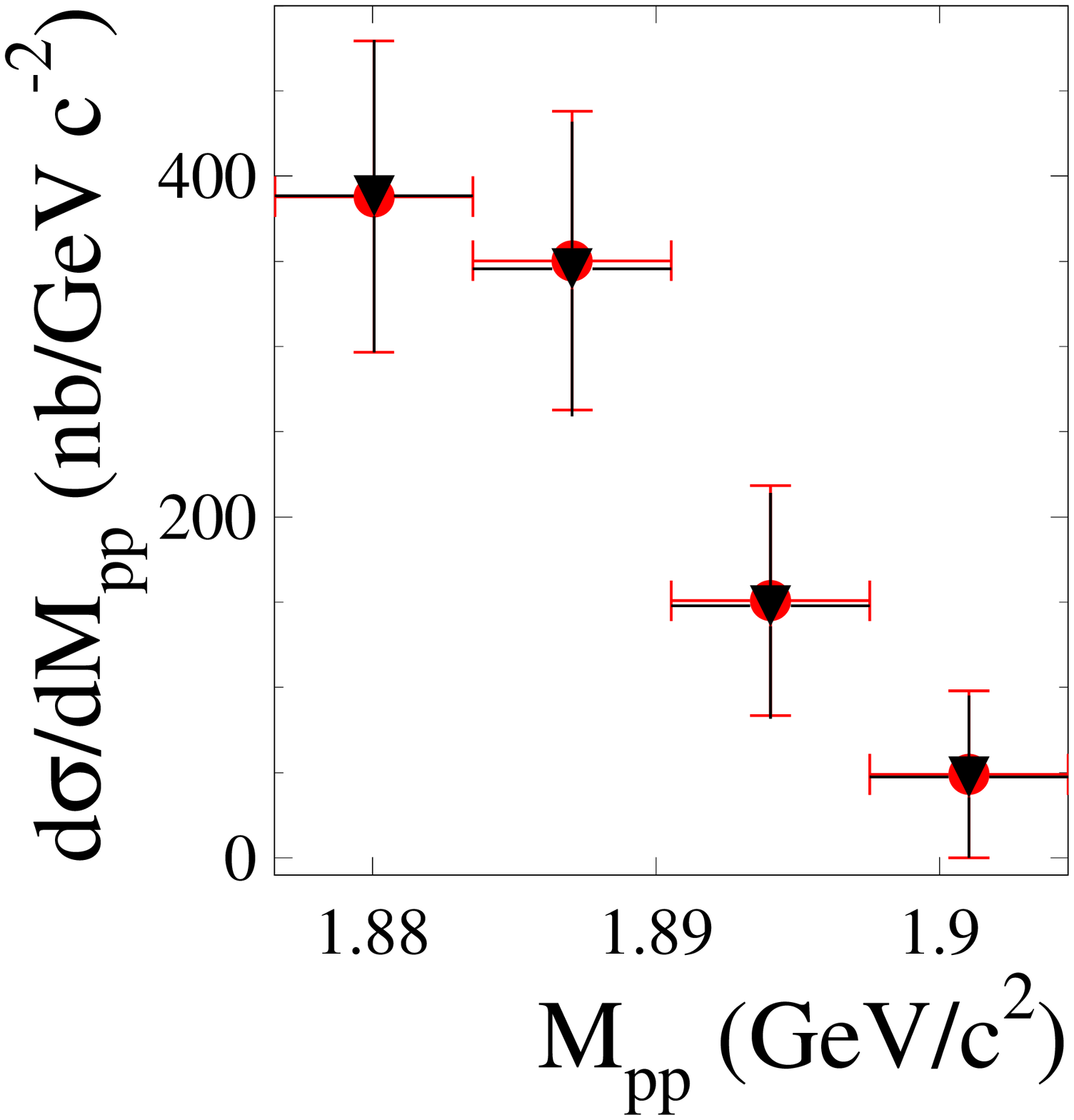}
\includegraphics[width=0.306\textwidth,angle=0]{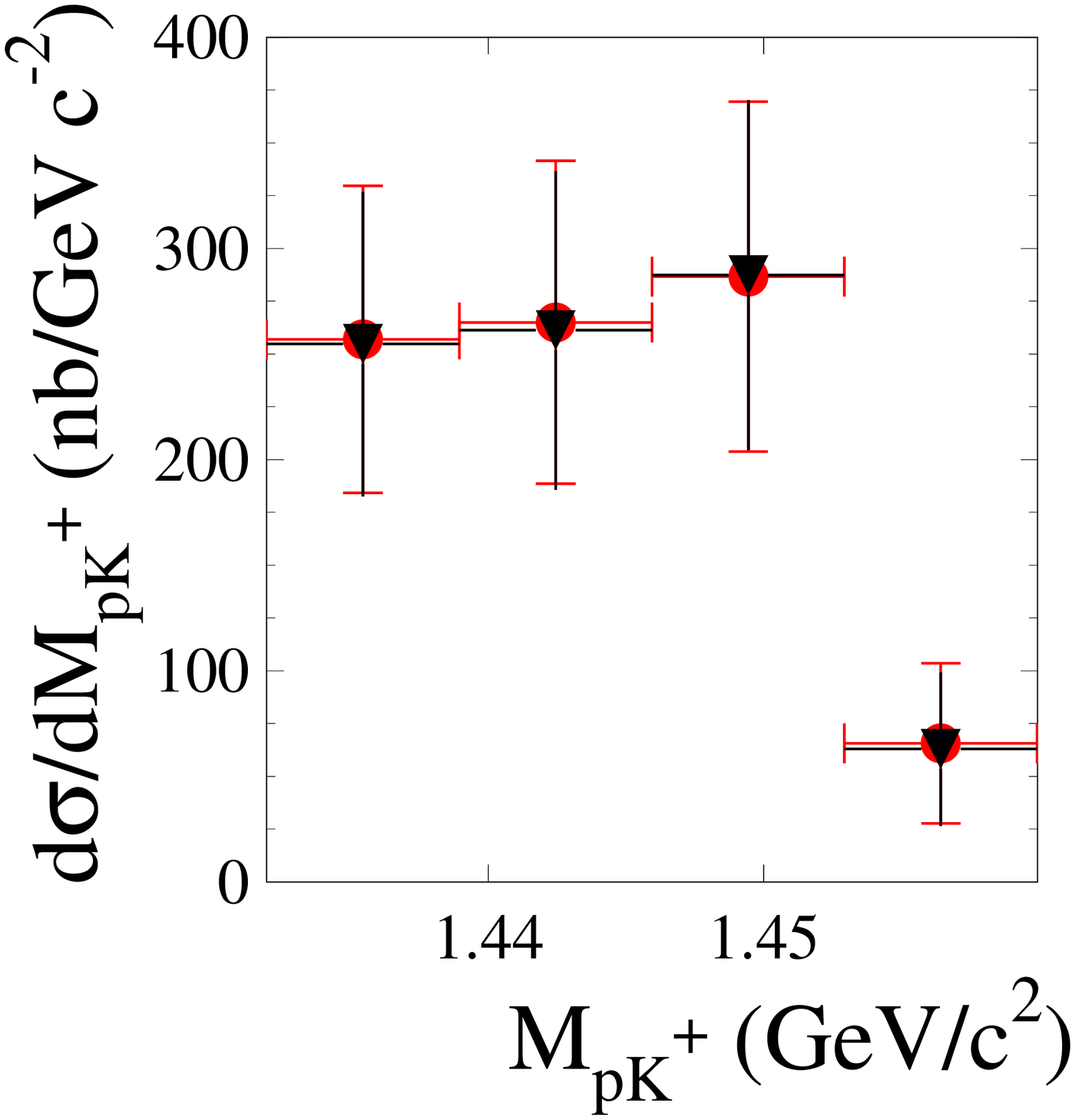}
\includegraphics[width=0.3\textwidth,angle=0]{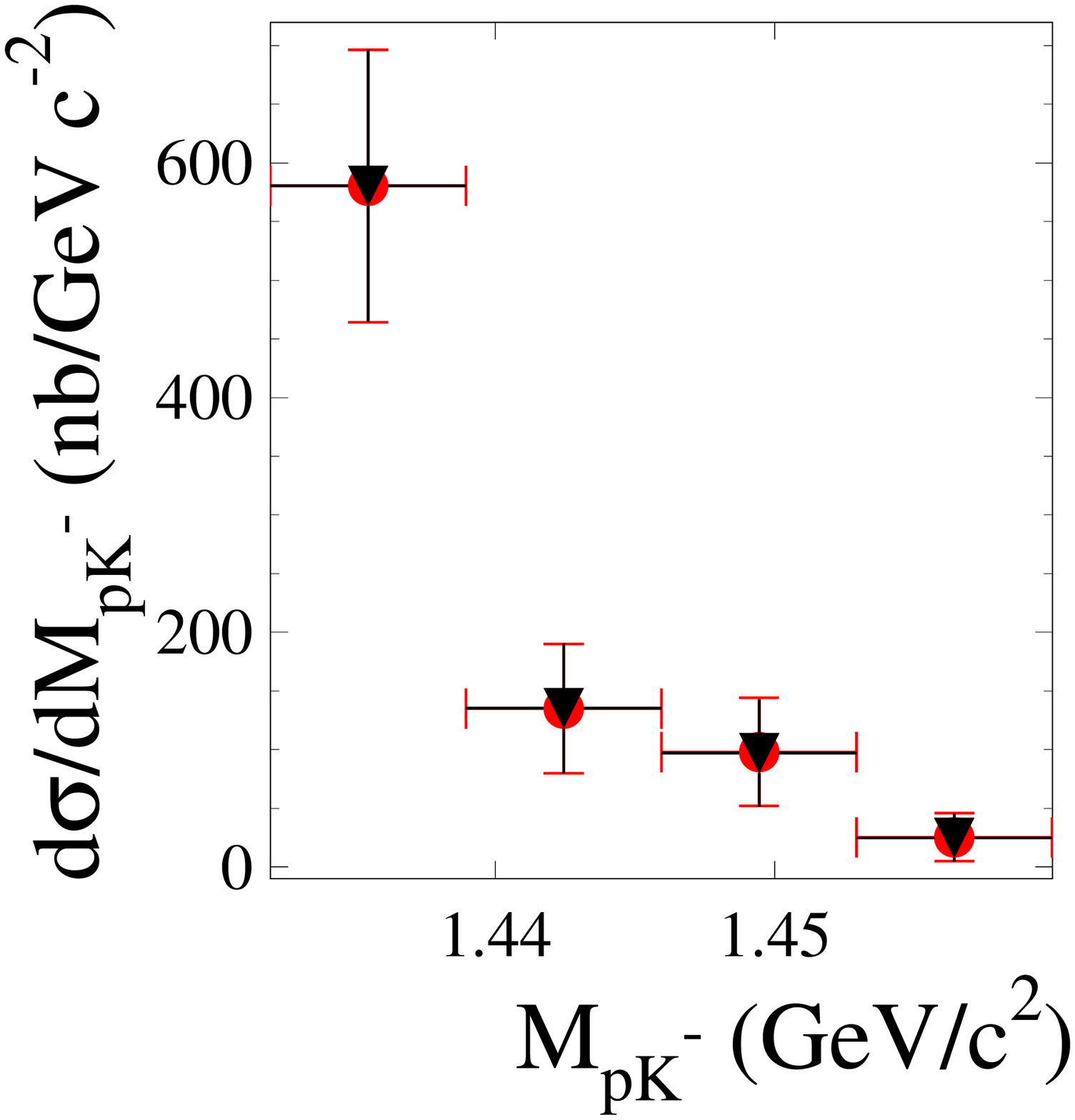}
\includegraphics[width=0.306\textwidth,angle=0]{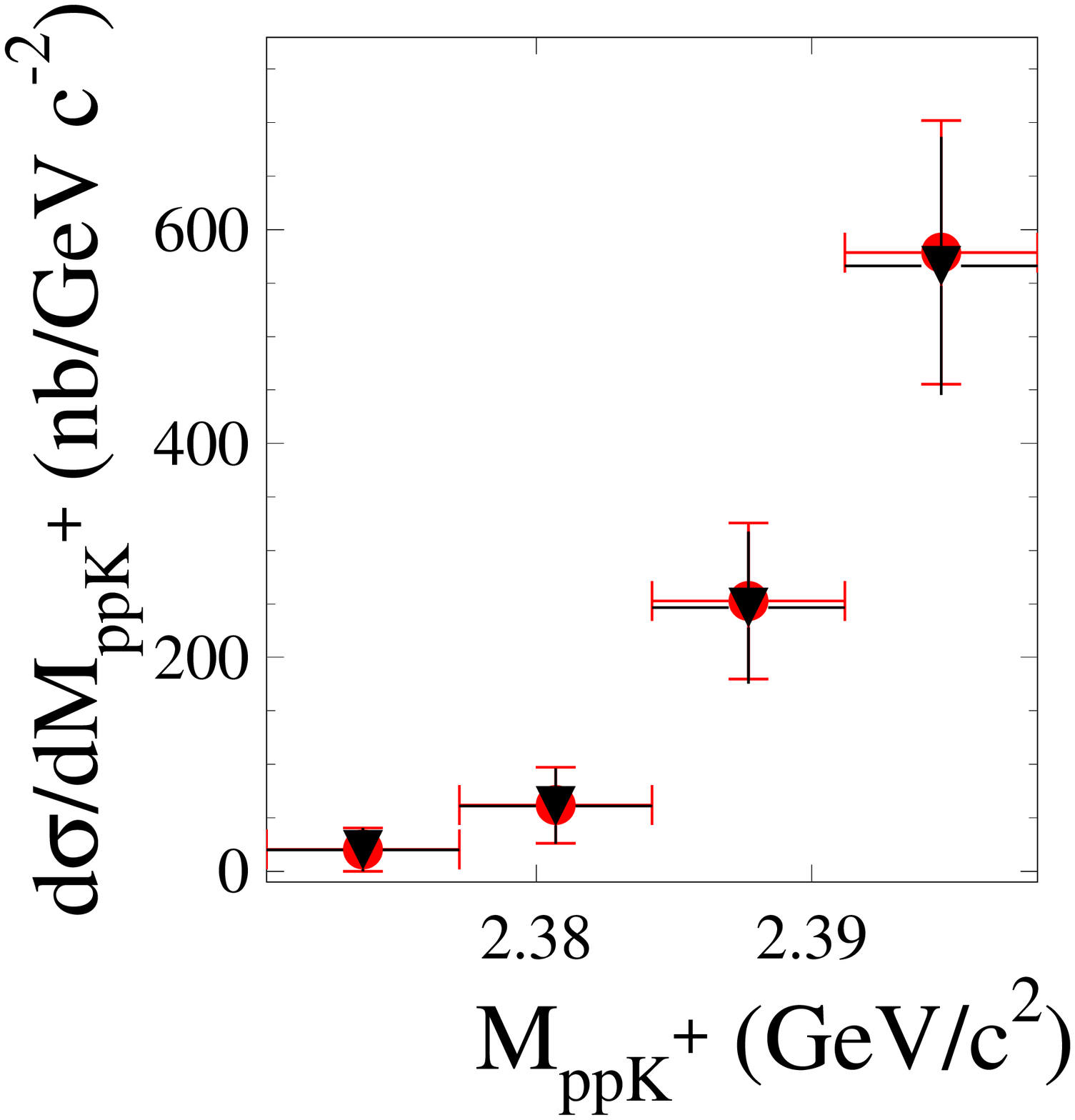}
\includegraphics[width=0.3\textwidth,angle=0]{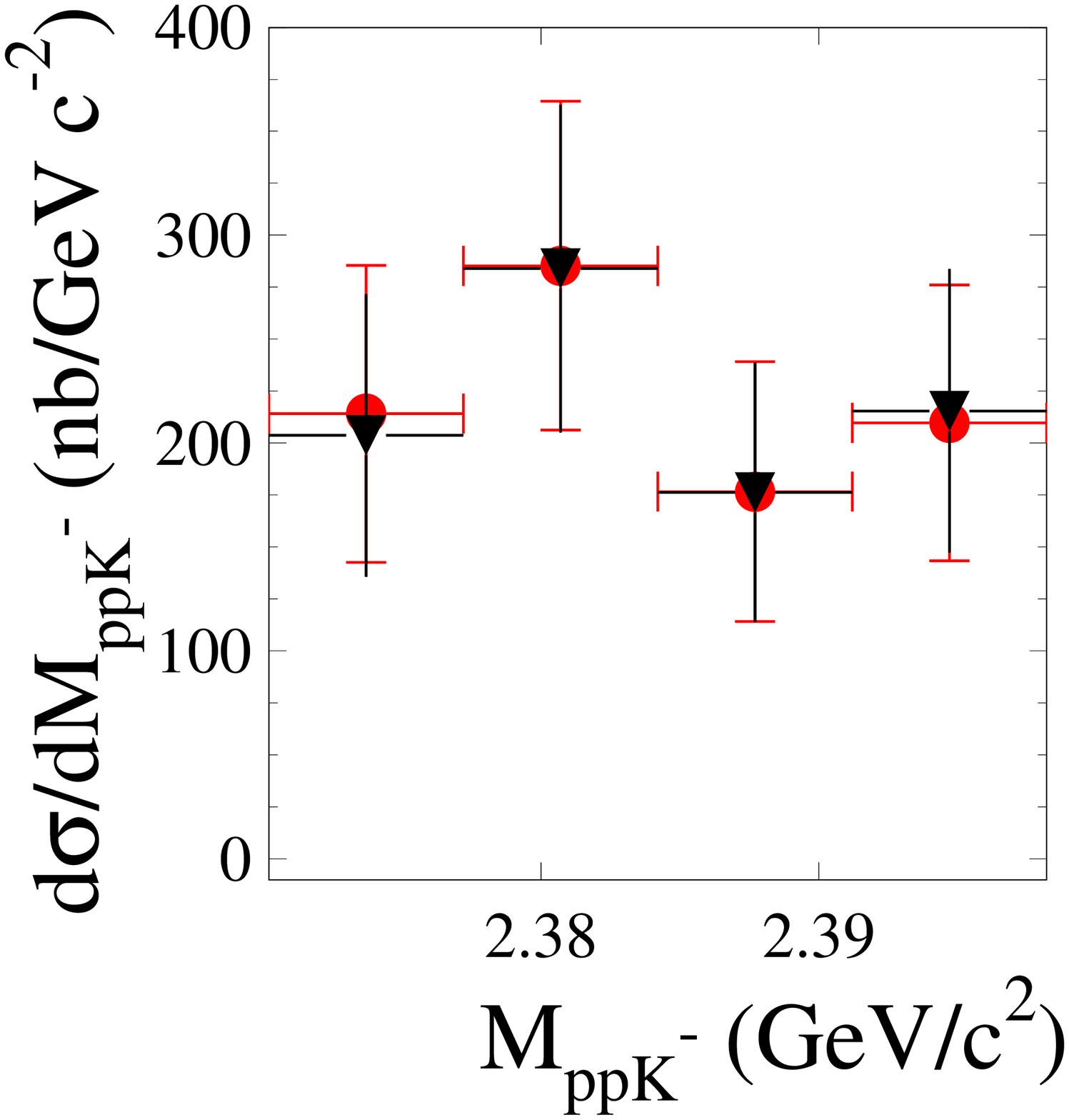}
\caption{The differential cross sections for the $pp\rightarrow ppK^{+}K^{-}$ reaction
at Q~=~28~MeV. Circles denote spectra where the acceptance was determined taking 
into account only the $pp$--FSI, and triangles  
denote results where additionally the $pK^{-}$--FSI was taken into account in 
the acceptance calculations. They are hardly distinguishable.}
\label{diff28}
\end{figure}

The authors of publication~\cite{anke} pointed out that the observed enhancement of the total 
cross section near threshold may be, at least partially, due to the neglect of the 
$pK^{-}$--FSI in the calculations of the COSY-11 acceptance. 
As a consequence the obtained total cross sections might decrease, if the interaction would have 
been taken into account during the analysis of the experimental data. 
This suggestion encouraged us to check quantitatively the influence 
of the interaction in the $pK^{-}$ subsystem on the acceptance of 
the detection setup. 
To this end we derived the distributions of the differential cross section 
for data at both excess energies assuming that the acceptance depends only on the $pp$--FSI.
Then we calculated the acceptance with inclusion of the $pK^{-}$--FSI and derived analogous
distributions.
\begin{figure}[h]
\centering 
\includegraphics[width=0.3\textwidth,angle=0]{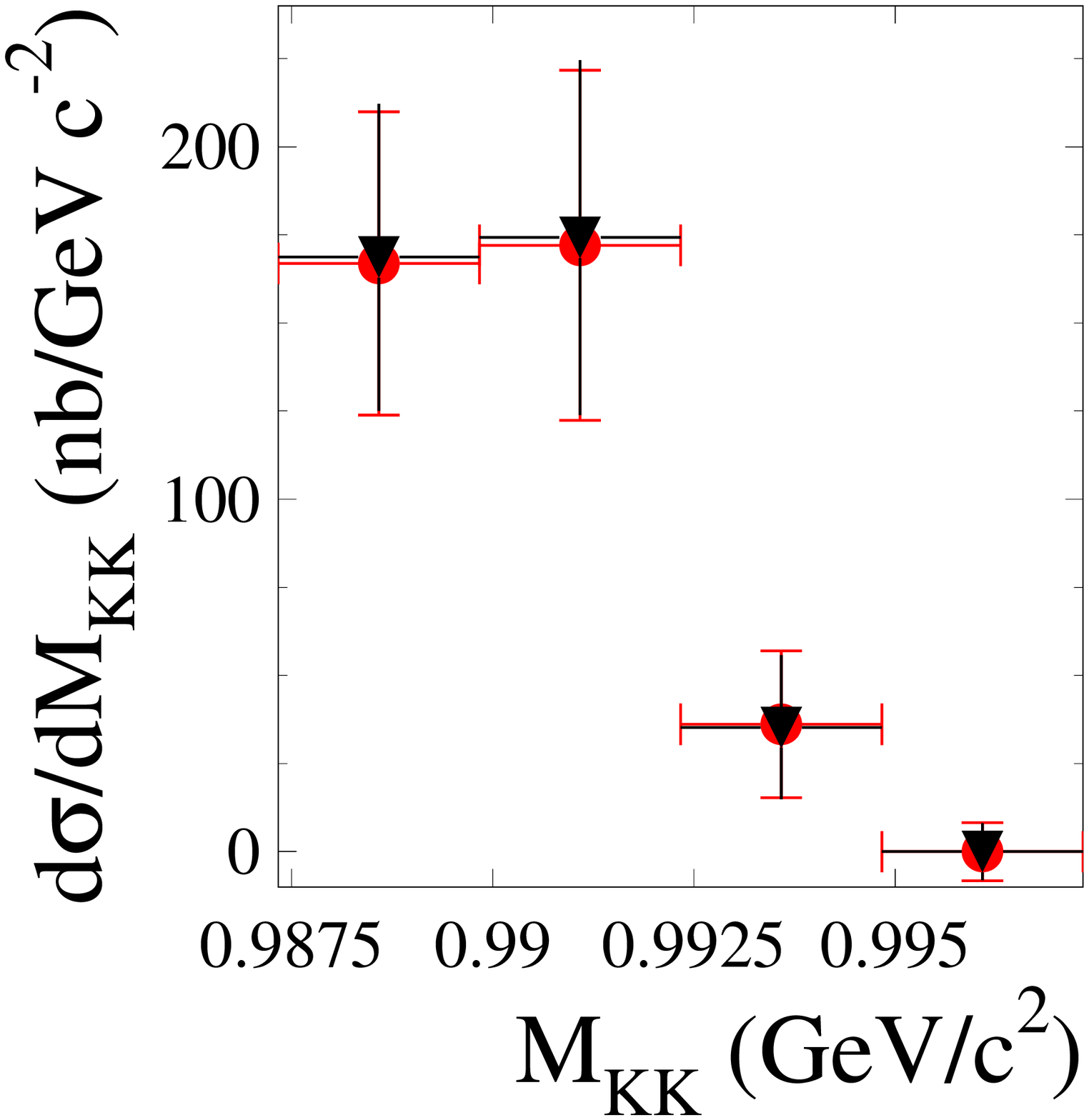}
\includegraphics[width=0.298\textwidth,angle=0]{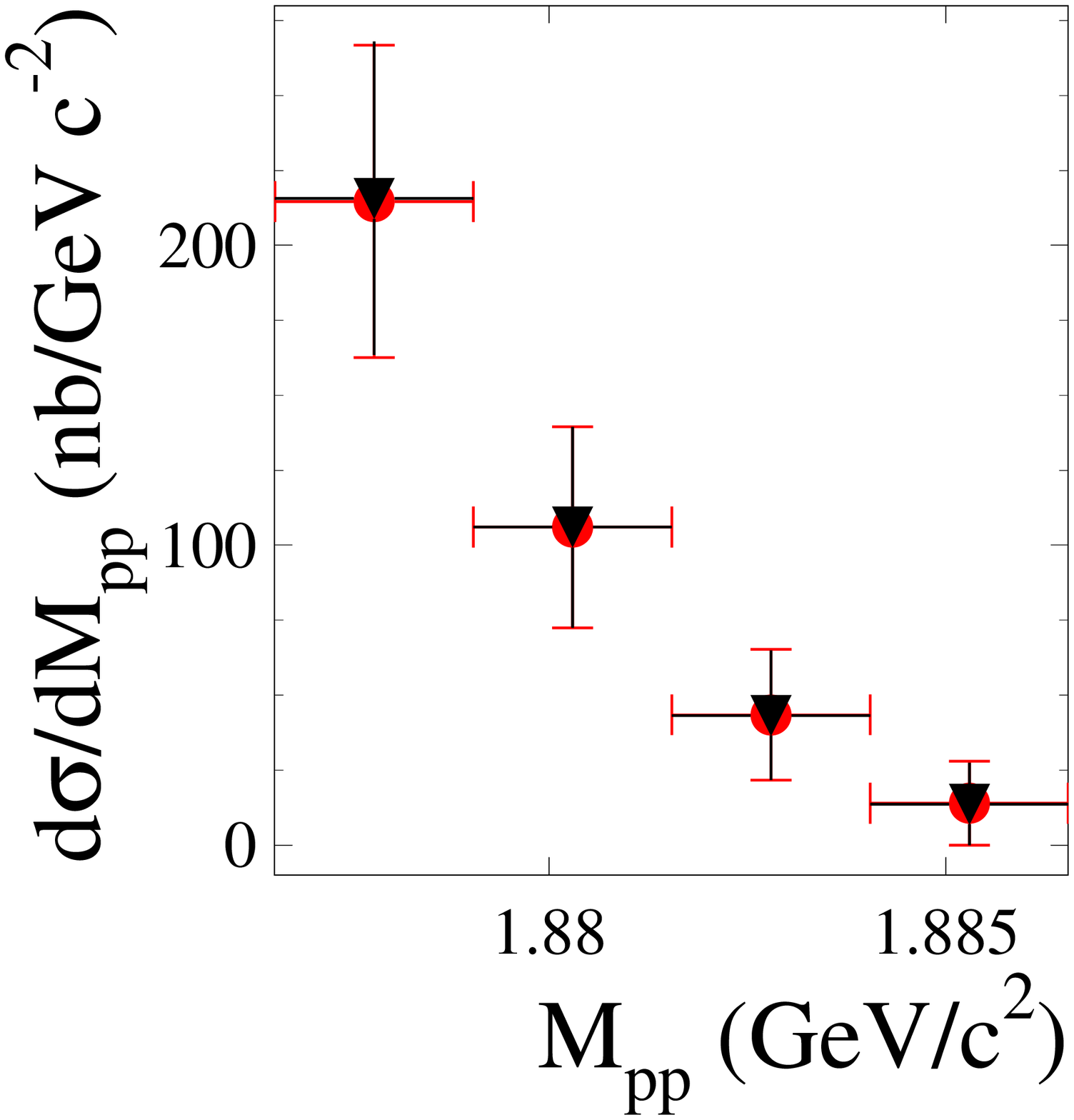}
\includegraphics[width=0.308\textwidth,angle=0]{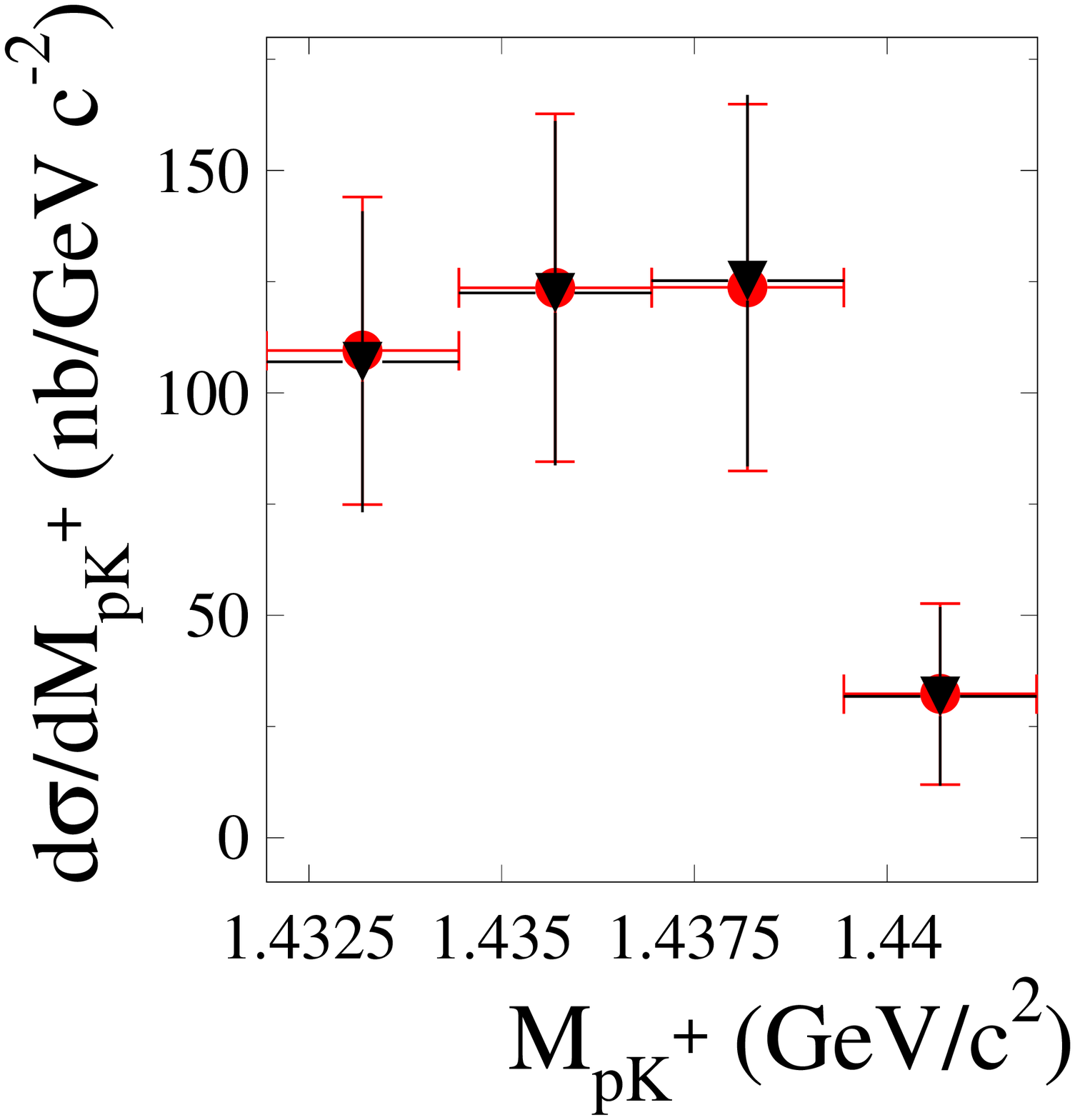}
\includegraphics[width=0.3\textwidth,angle=0]{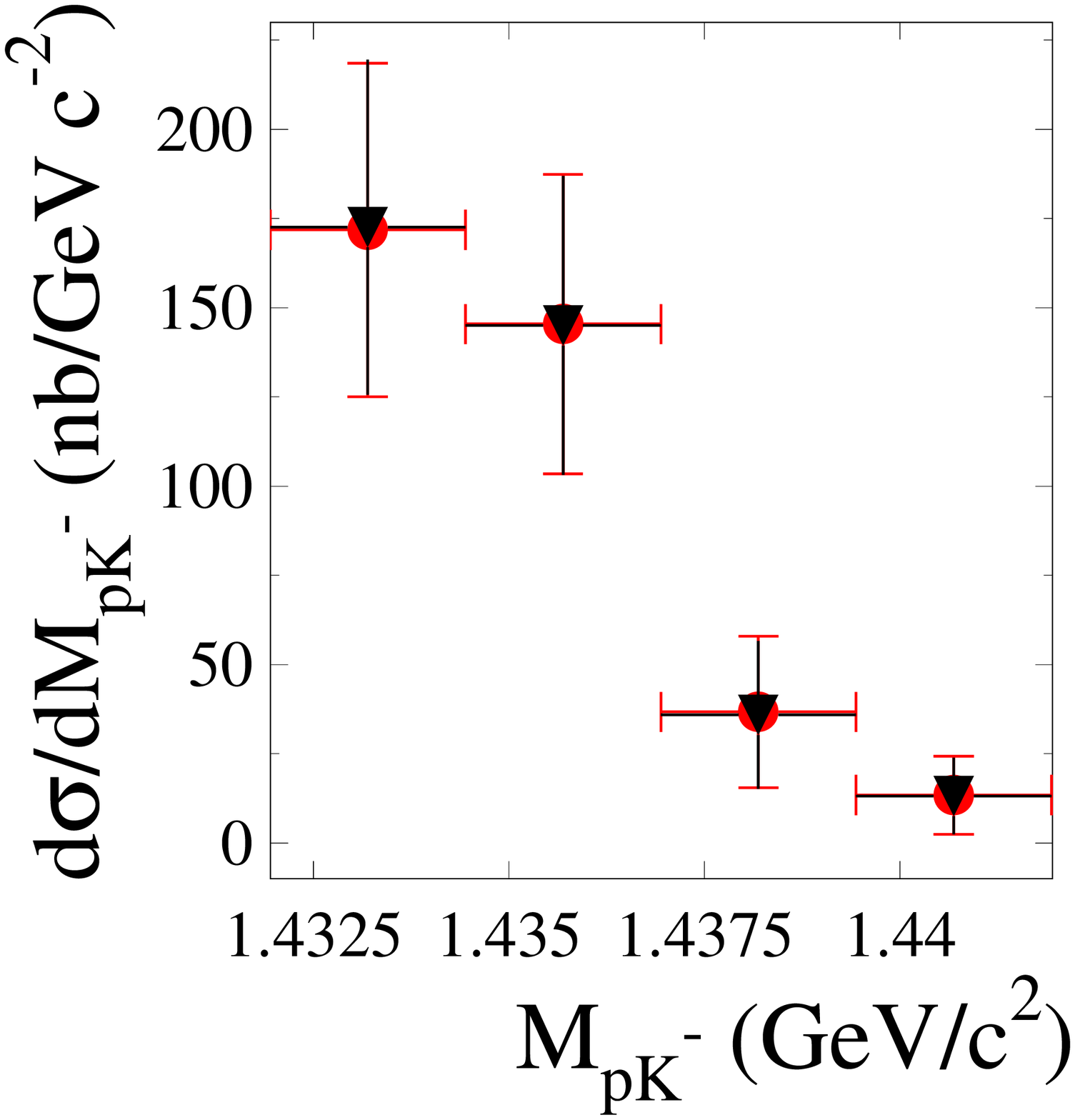}
\includegraphics[width=0.306\textwidth,angle=0]{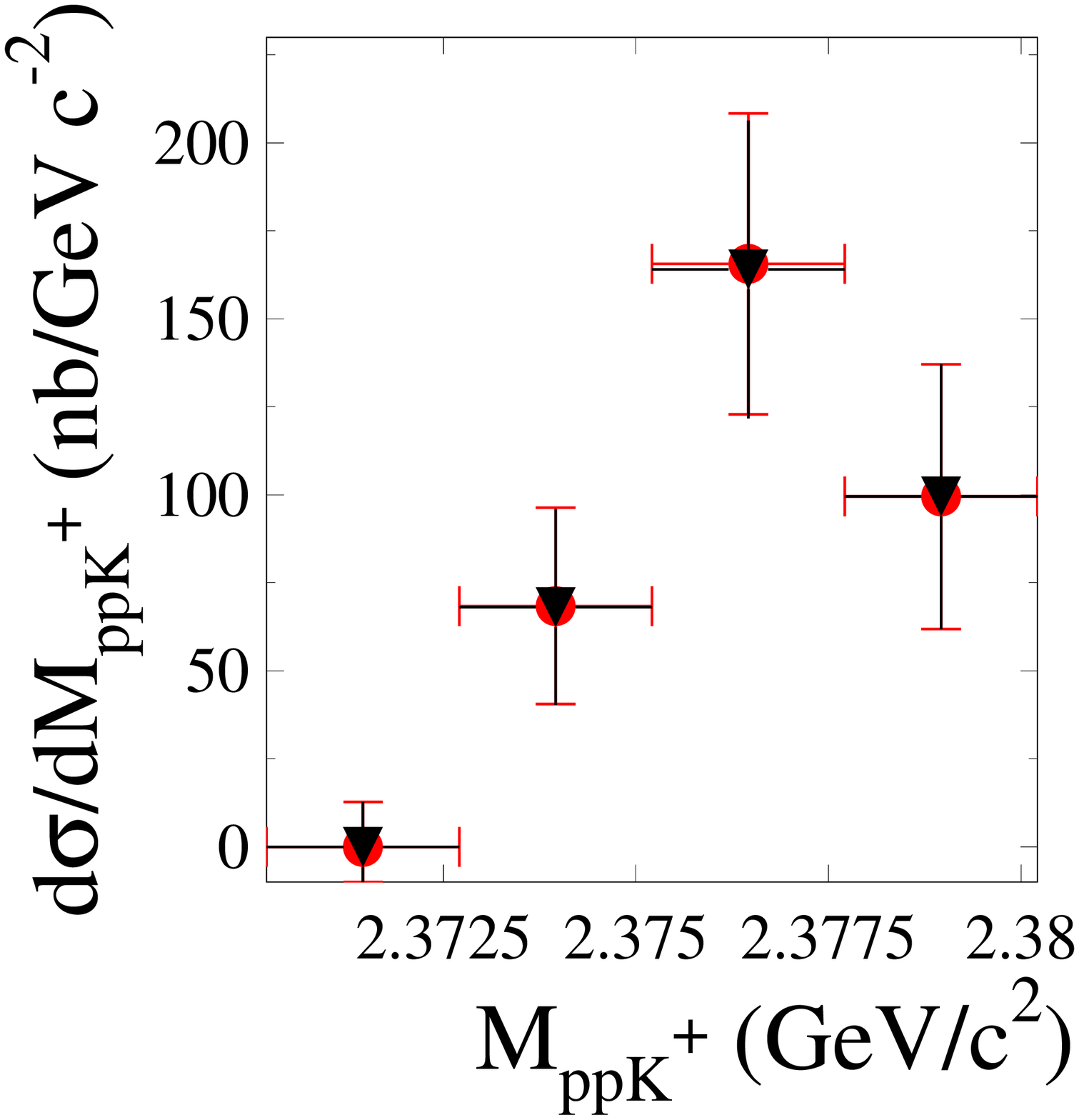}
\includegraphics[width=0.306\textwidth,angle=0]{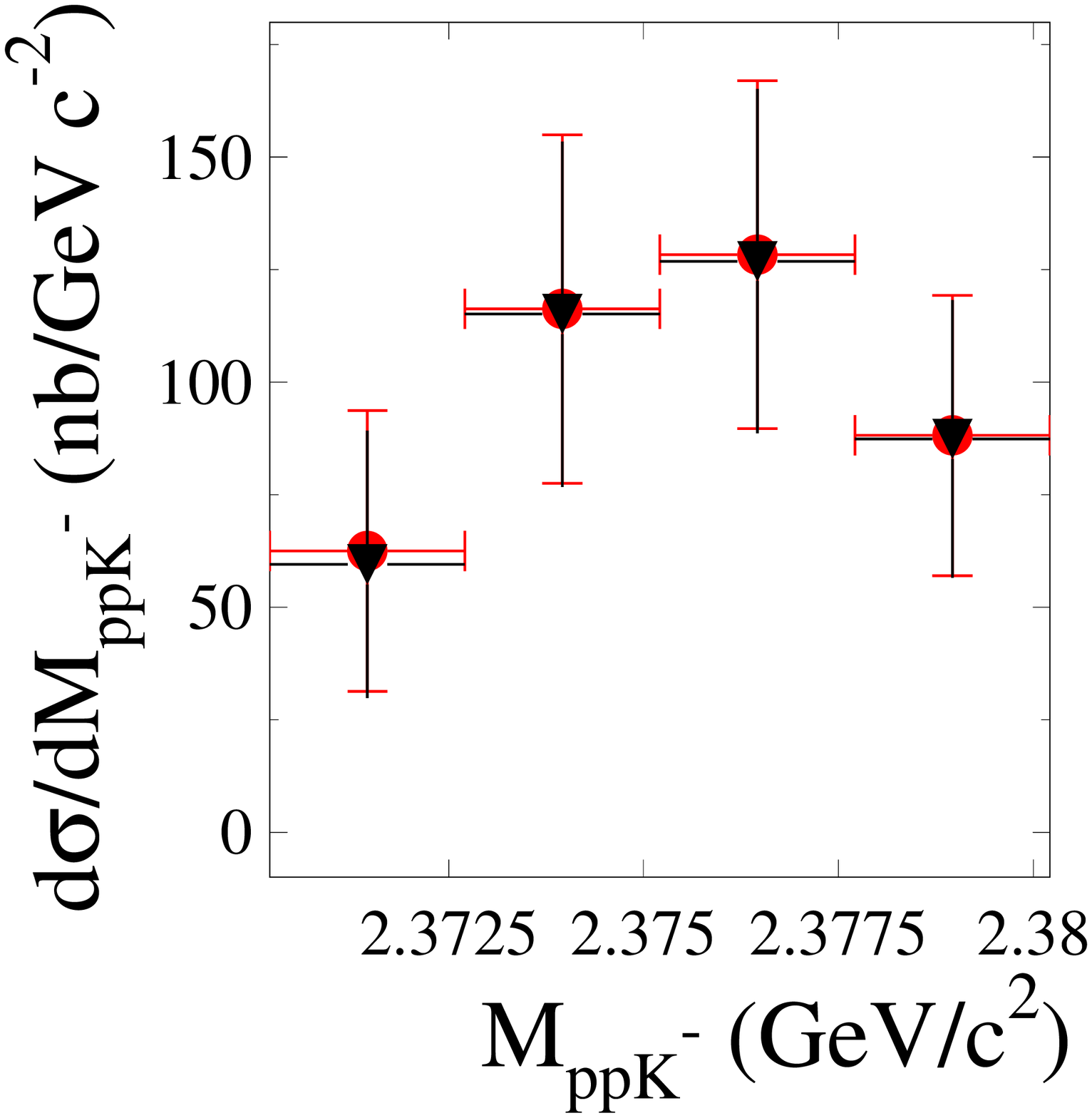}
\caption{The differential cross sections for the $pp\rightarrow ppK^{+}K^{-}$ reaction
at Q~=~10~MeV. Circles denote spectra where the acceptance was determined taking 
into account only the $pp$--FSI, and triangles  
denote results where additionally the $pK^{-}$--FSI was taken into account in 
the acceptance calculations. They are hardly distinguishable.}
\label{diff10}
\end{figure}
The results are presented in Fig.~\ref{diff10} for data at 
Q~=~10~MeV and in Fig.~\ref{diff28} for Q~=~28~MeV.
As one can see, distributions obtained under both assumptions are almost identical, which shows 
that the acceptance of the COSY-11 detection setup is only very weakly sensitive to the 
interaction between $K^{-}$ and protons. \\
The spectra shown in Figs.~\ref{diff10} and~\ref{diff28} provide an additional 
information to the total cross sections published previously~\cite{Winter}, 
where the values of the cross sections were determined using the total 
number of events identified as the $pp\rightarrow ppK^{+}K^{-}$ reaction and 
the total acceptance of the COSY-11 detector system. The acceptance was 
calculated including the $pp$--FSI described by the on shell proton--proton 
scattering amplitude. Now after the determination of the absolute values 
for the differential distributions one can calculate the total cross 
sections in a less model dependent manner regardless of the assumption 
of the $pp$--FSI.
The cross sections, calculated for both excess energies as a integral of the $M_{pp}$
distribution derived with inclusion of the $pK^{-}$--FSI in the acceptance calculations:
 \begin{displaymath}
\sigma_{tot}~=~\int\frac{d\sigma}{dM_{pp}}dM_{pp}~,
\end{displaymath}
amount to $\sigma_{tot}~=~(0.95~\pm~0.17)~$nb for measurement at Q~=~10~MeV and
$\sigma_{tot}~=~(6.5~\pm~1.1)~$nb for Q~=~28~MeV.
These results are larger than the previously obtained total cross sections by about $20~\%$ 
for Q~=~10~MeV and 50~\% for Q~=~28~MeV, which strengthen the confidence to the 
observed enhancement at threshold.
However, the total cross sections obtained in these two different analyses 
are statistically consistent.
\begin{figure}[h]
\centering
\includegraphics[width=0.35\textwidth,angle=0]{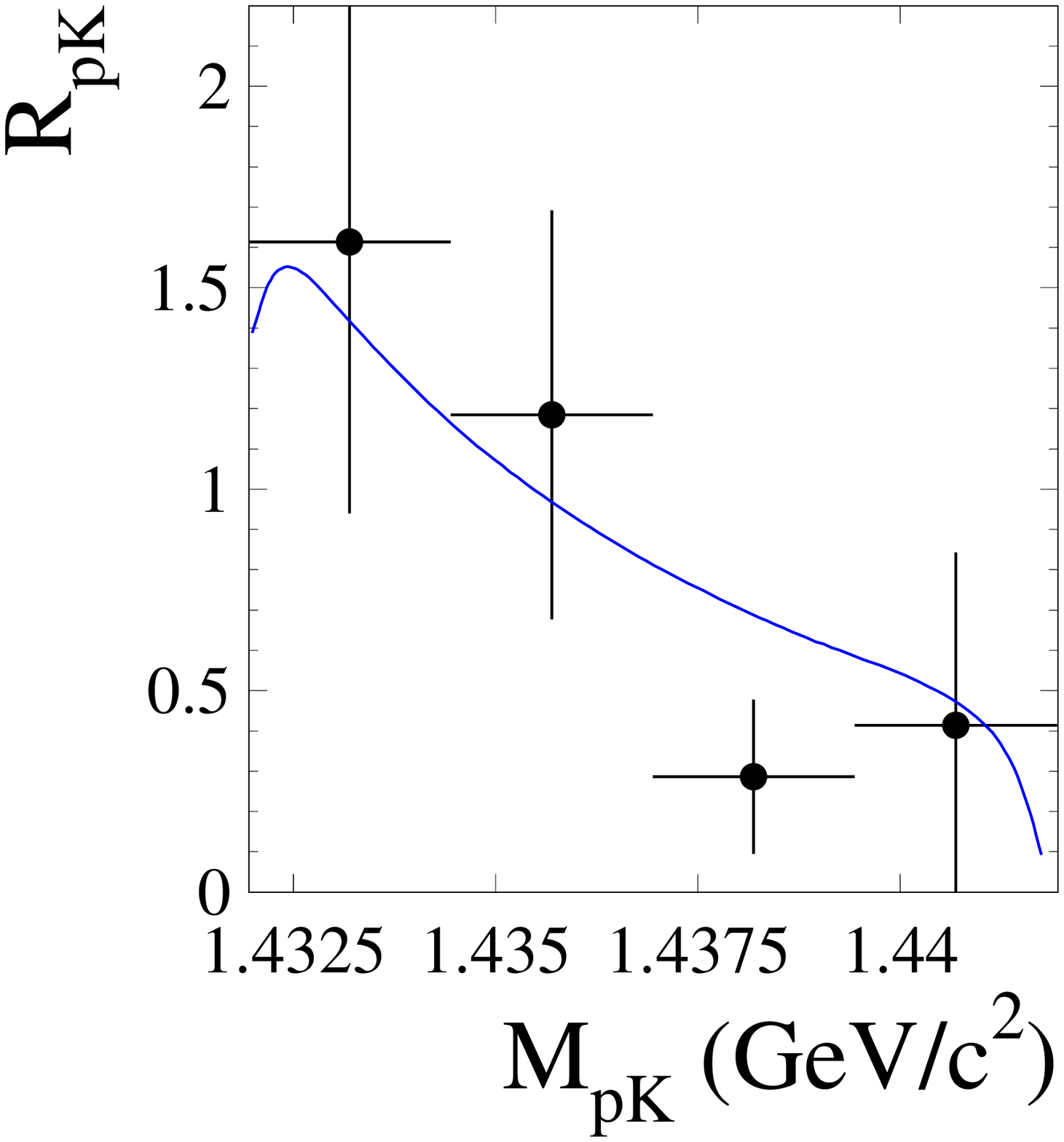}
\includegraphics[width=0.362\textwidth,angle=0]{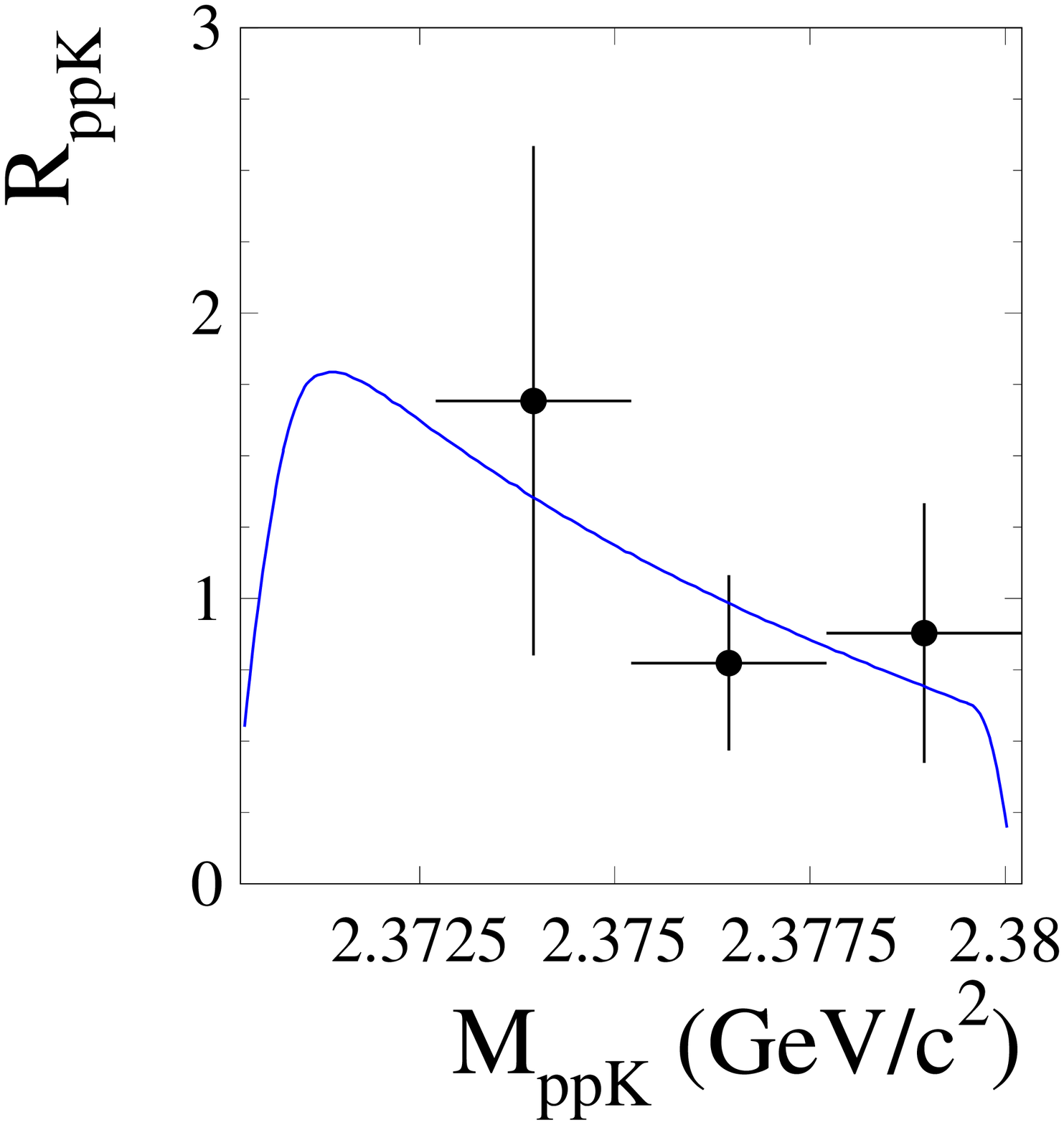}
\includegraphics[width=0.35\textwidth,angle=0]{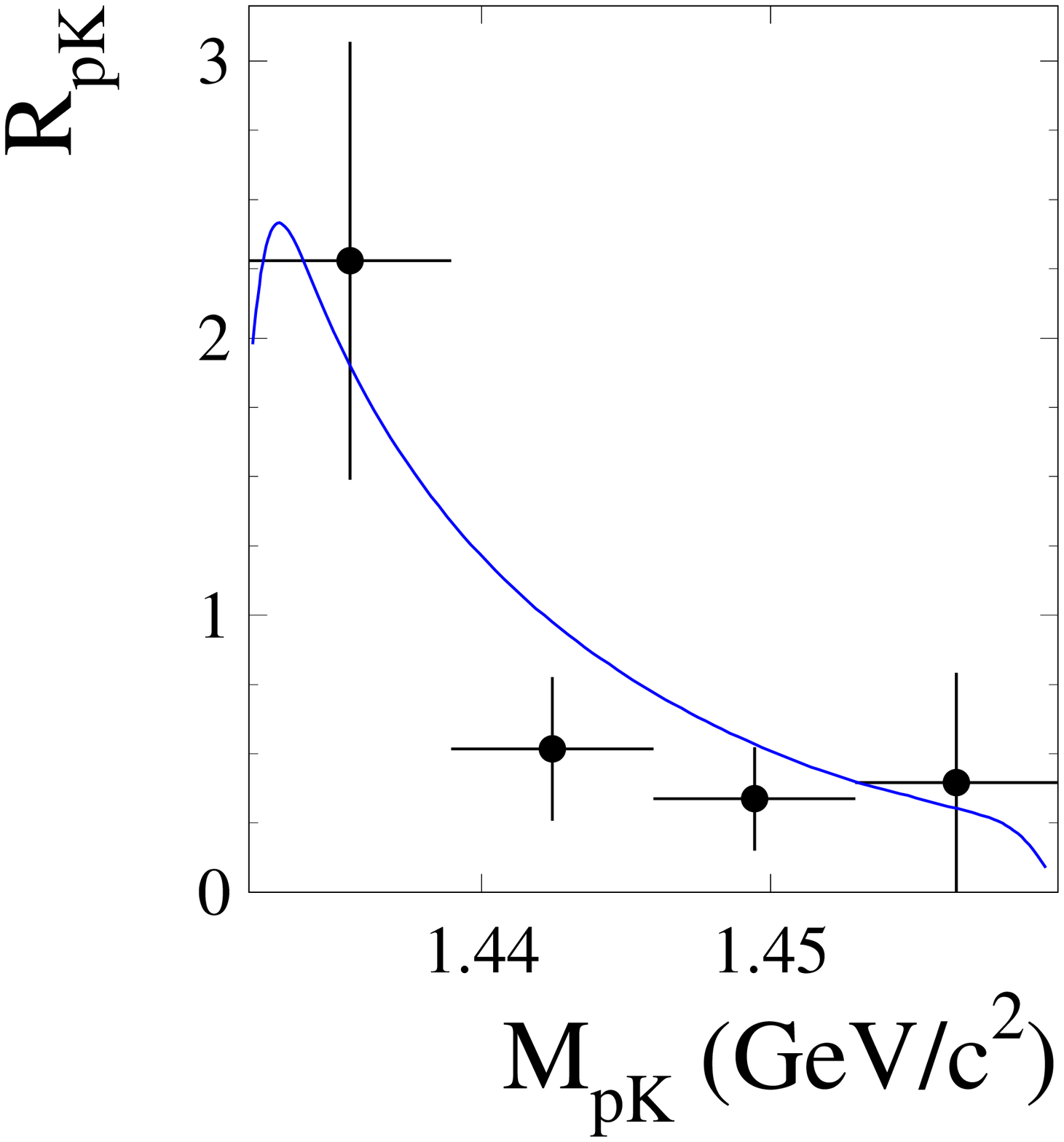}
\includegraphics[width=0.35\textwidth,angle=0]{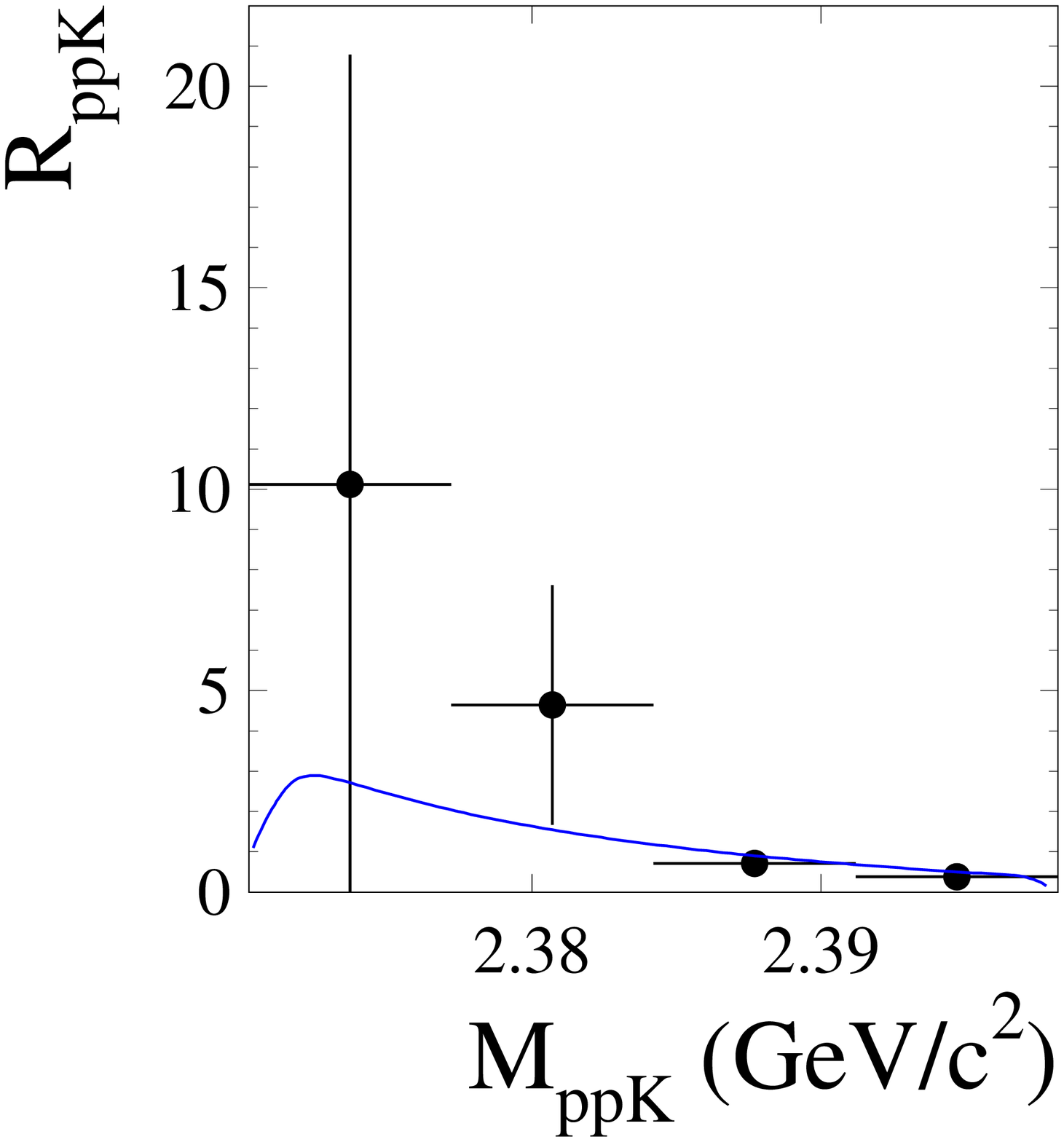}
\caption{The distributions of ratios $R_{pK}$ and $R_{ppK}$ for data at Q~=~10~MeV (upper panel) 
and Q~=~28~MeV (lower panel). Solid curves represent theoretical expectations calculated
taking into account $pp$ and $pK^{-}$ final state interaction.}
\label{ratio}
\end{figure}
The determination of the absolute values for the differential cross sections 
permitted us to establish the absolute values for the following ratios at the 
close to threshold region:
\begin{displaymath}
R_{pK}~=~\frac{d\sigma/dM_{pK^{-}}}{d\sigma/dM_{pK^{+}}}~,
\end{displaymath}
\begin{displaymath}
R_{ppK}~=~\frac{d\sigma/dM_{ppK^{-}}}{d\sigma/dM_{ppK^{+}}}~.
\end{displaymath} 
If $pK^+$ and $pK^-$ interactions were the same, the distribution of $R_{pK}$ as well as 
$R_{ppK}$ should be flat and equal to unity. But as one can see in Fig.~\ref{ratio} 
and as presented already in the previous publication by COSY-11~\cite{Winter} and 
ANKE~\cite{anke} 
$R_{pK}$ for 
both excess energies is far from constant and increases towards the 
lower $M_{pK}$ invariant masses. This effect might be connected with the influence of the $pK^{-}$ 
final state interaction. Similarly the distributions of $R_{ppK}$ differs from 
expectations assuming only interaction in the $pp$ system. 
This is a confirmation of effects found also by the ANKE 
collaboration at higher excess energies~\cite{anke}. 
As one can see in Fig.~\ref{ratio} simulations taking into account the 
$pK^{-}$ final state interaction with the scattering length determined 
by the ANKE group for the data at significantly higher excess energies reproduce 
very well the distributions of $R_{pK}$ and $R_{ppK}$ near the threshold.
The results presented by the curves in Fig.~\ref{ratio} were determined assuming 
that the $pK^-$ scattering length amounts to: $a_{pK^{-}}$~=~(0~+~1.5$i$)~fm~\cite{anke}
\footnote
{In  this calculations we used the following parametrization of the proton-proton scattering amplitude:
\begin{displaymath}
F_{pp} = 
  \frac{e^{-i\delta_{pp}({^{1}\mbox{\scriptsize S}_{0}})} \cdot 
        \sin{\delta_{pp}({^{1}\mbox{S}_0})}}
       {C \cdot \mbox{q}}~,
\end{displaymath}
where $C$ stands for the square root of the Coulomb penetration factor~\cite{pp-FSI}. 
The parameter $\delta_{pp}({^{1}\mbox{S}_0})$ denotes the phase-shift calculated according
 to the modified Cini-Fubini-Stanghellini formula with the Wong-Noyes Coulomb 
correction~\cite{noyes995,naisse506,noyes465}. A more detailed description of this parametrization 
can be found in references~\cite{pp-FSI,noyes995,naisse506,noyes465,habilitacja}.
}.
\section{Conclusions}
We concluded, that a reanalysis of the COSY-11 data with the inclusion of the $pK^{-}$ interaction 
did not change significantly the shape of the previously determined differential distributions of the 
cross section.  
Moreover, the determination of the total 
cross sections from the differential $M_{pp}$ distributions even increased the observed 
enhancement at threshold. 
Regarding  the comparison of the interactions in the $pK^{-}$, $pK^{+}$, $ppK^{-}$ and 
$ppK^{+}$ subsystems, the absolute ratios determined from COSY-11 data measured at 
Q~=~10~MeV and Q~=~28~MeV are consistent with the predictions based on parametrisation 
introduced in reference~\cite{anke} and on the values of the scattering length $a_{K^-p}$ 
extracted from the ANKE data at higher excess energies~\cite{anke}.
\section{Acknowledgements}
The work was
supported by the
European Community-Research Infrastructure Activity
under the FP6 program (Hadron Physics,RII3-CT-2004-506078), by 
the German Research Foundation (DFG), by
the Polish Ministry of Science and Higher Education through grants
No. 3240/H03/2006/31  and 1202/DFG/2007/03,
and by the FFE grants from the Research Center J{\"u}lich.

\end{document}